\documentclass[iop,apjl]{emulateapj}   

\pdfoutput=1

\usepackage{graphicx}
\usepackage{epsfig}
\usepackage{amsmath}
\usepackage{subfigure} 
\usepackage{hyperref}
\usepackage{amssymb}
\usepackage{apjfonts}
\usepackage{latexsym}
\usepackage{color}
\usepackage[usenames,dvipsnames,svgnames,table]{xcolor}
\usepackage{hyperref}
\usepackage[all]{hypcap}    
\hypersetup{
    colorlinks,
    citecolor=blue,
    filecolor=blue,
    linkcolor=blue,
    urlcolor=blue
}

\def\kms{\ifmmode{~{\rm km~s^{-1}}}\else{~km s$^{-1}$}\fi}
\def\cc{\ifmmode{~{\rm cm^{-3}}}\else{~cm$^{-3}$}\fi}
\def\fesc{\ifmmode{{f_{esc}}}\else{$f_{\rm esc}$}\fi}
\def\fstar{\ifmmode{{f_\star}}\else{$f_\star$}\fi}
\def\lsim{\lower0.3em\hbox{$\,\buildrel <\over\sim\,$}}
\def\gsim{\lower0.3em\hbox{$\,\buildrel >\over\sim\,$}}

\def\enzo{{\sc Enzo}}
\def\moray{{\sc Moray}}
\def\yt{{\sc yt}}
\def\Ms{\ifmmode{~{\rm M_\odot}}\else{M$_\odot$}\fi}
\def\Zs{\ifmmode{~{\rm Z_\odot}}\else{Z$_\odot$}\fi}
\def\h2{H$_2$}

\def\nat{Nature}
\def\apj{ApJ}
\def\aj{AJ}
\def\apjl{ApJL}
\def\apjs{ApJS}
\def\mnras{MNRAS}

\newcommand{\hii}{H {\sc ii}}

\shorttitle{Probing the High-z UV Luminosity Function} 
\shortauthors{B.W. O'Shea et al.}

\begin{document}\title{Probing The Ultraviolet Luminosity Function of
  the Earliest Galaxies with the Renaissance Simulations}

\author{
  Brian W. O'Shea\altaffilmark{1,2,3},
  John H. Wise\altaffilmark{4},
  Hao Xu\altaffilmark{5}, and
  Michael L. Norman\altaffilmark{5}}

\affil{$^{1}${Department of Physics and Astronomy, Michigan State
    University, East Lansing, MI 48824, USA;
    \href{mailto:oshea@msu.edu}{oshea@msu.edu}}}

\affil{$^{2}${Department of Computational Mathematics, Science and Engineering, East
    Lansing, MI 48824, USA}}

\affil{$^{3}${JINA: Joint Institute for Nuclear Astrophysics}}

\affil{$^{4}${Center for Relativistic Astrophysics, School of Physics,
    Georgia Institute of Technology, 837 State Street, Atlanta, GA
    30332; \href{mailto:jwise@gatech.edu}{jwise@gatech.edu}}}

\affil{$^{5}${CASS, University of California, San Diego, 9500 Gilman
    Drive, La Jolla, CA 92093;
    \href{mailto:hxu@ucsd.edu}{hxu@ucsd.edu},
    \href{mailto:mlnorman@ucsd.edu}{mlnorman@ucsd.edu}}}

\label{firstpage}

\begin{abstract}
In this paper, we present the first results from the
\textit{Renaissance Simulations}, a suite of extremely high-resolution
and physics-rich AMR calculations of high redshift galaxy formation
performed on the Blue Waters supercomputer.  These simulations contain
hundreds of well-resolved galaxies at $z \sim 25-8$, and make several
novel, testable predictions.  Most critically, we show that the
ultraviolet luminosity function of our simulated galaxies is
consistent with observations of high-z galaxy populations at the
bright end of the luminosity function (M$_{1600} \leq -17$), but at
lower luminosities is essentially flat rather than rising steeply, as
has been inferred by Schechter function fits to high-z observations,
and has a clearly-defined lower limit in UV luminosity.  This behavior
of the luminosity function is due to two factors: (i) the strong
dependence of the star formation rate on halo virial mass in our
simulated galaxy population, with lower-mass halos having
systematically lower star formation rates and thus lower UV
luminosities; and (ii) the fact that halos with virial masses below
$\simeq 2 \times 10^8$~M$_\odot$ do not universally contain stars,
with the fraction of halos containing stars dropping to zero at
$\simeq 7 \times 10^6$~M$_\odot$.  Finally, we show that the brightest
of our simulated galaxies may be visible to current and future
ultra-deep space-based surveys, particularly if lensed regions are
chosen for observation.

\end{abstract}


\section{Introduction}
\label{sec:introduction}

The most fundamental characteristics of the earliest galaxies are
challenging to determine directly.  These galaxies lie at the edge of
observability, or beyond, for even the largest current ground- and
space-based telescopes, and inferences based on local stellar
populations (from, e.g, the Milky Way's stellar halo) are uncertain at
best.  In particular, it is difficult to determine the characteristics
of the luminosity function of high redshift galaxies, which directly
influences how reionization proceeds in the early universe.  A recent
example of this are efforts to use the Hubble Ultra Deep Field and its
extensions
\citep{2006AJ....132.1729B,2014arXiv1410.5439F,2013ApJS..209....3K,2015ApJ...803...34B},
and, separately, lensed observations of high redshift galaxies from
the Hubble Frontier Fields
\citep{2015ApJ...800...18A,2015ApJ...799...12I},
to estimate the luminosity function for galaxies at $z \simeq 6-10$.
These observations succeeded in directly measuring the stellar
luminosity function for galaxies with luminosities within a few
magnitudes of L$^*$, but no deeper.

The work of \citet{2015ApJ...803...34B} and
\citet{2015ApJ...800...18A} provides only weak constraints on the
faint-end slope of the luminosity function, which has profound
theoretical implications
\citep[e.g.,][]{2010Natur.468...49R,2014MNRAS.438.2097F}. If the slope
of the luminosity function is very steep, this implies a vast number
of dim galaxies, unobservable by current instruments, that can produce
more than enough ionizing photons to complete and sustain reionization
with no additional sources (such as high-redshift black hole
populations) needed.  If, on the other hand, the slope of the
high-redshift luminosity function is shallow, there will be fewer
galaxies than one might expect, with too few ionizing photons and thus
the need for additional sources of ionizing radiation.

A second question related to the shape of the high-redshift luminosity
function is that of its ending point: how bright are the smallest
high-redshift galaxies?  Simulations of Population III stars and the
transition to metal-enriched star formation suggest that the smallest
halos to form stars have masses of around $10^6 - 10^7$~M$_\odot$
\citep{2007ApJ...654...66O,2008MNRAS.388...26J,2009ApJ...691..441S,2012ApJ...745...50W,2013ApJ...773..108C},
with correspondingly few stars and low luminosities.  These
simulations' predictive capabilities, however, suffer from challenges
relating to small-number and small-volume statistics.

An additional issue relating to the study of high redshift galaxies
relates to cosmic variance.  The James Webb Space Telescope will
certainly be able to see substantial numbers of galaxies at $z \ga
10$, but has a small field of view \citep{2006SSRv..123..485G}.  As a
result, interpretation of any JWST survey must by necessity take into
account cosmic variance \citep[e.g.,][]{2008ApJ...676..767T,
2011ApJ...731..113M}.  While this can be done by using many small,
widely spaced fields instead of one larger field
\citep[][]{2013MNRAS.428L...6S}, this may be undesirable, and thus a
deeper understanding based on theory may be necessary to correctly
interpret high-z survey results.

In this paper, we address these pressing issues regarding the high
redshift galaxy luminosity function using the \textit{Renaissance
Simulations} -- a trio of physics-rich simulations of high redshift
galaxy formation that resolve several hundred galaxies apiece, and
which explore a range of cosmic environments at $z \geq 8$.  Using
these simulations (described in Section~\ref{sec:methods}), we show
results (in Section~\ref{sec:results}) that predict the high redshift
galaxy luminosity function will flatten at magnitudes that can be
probed by the next generation of telescopes. 
We discuss the implications of this work in
Section~\ref{sec:conclusions}.

\section{The Renaissance Simulations}
\label{sec:methods}

The simulations were carried out using 
\enzo\footnote{\url{http://enzo-project.org/}} 
\citep{EnzoMethodPaper}, an open-source adaptive mesh
refinement code that has been extensively used for simulating
cosmological structures, and in particular high-redshift structure
formation
\citep[e.g.,][]{2002Sci...295...93A,2007ApJ...654...66O,2009Sci...325..601T,2012MNRAS.427..311W,2012ApJ...745...50W,2013ApJ...773...83X,2014MNRAS.442.2560W}.
Notably, H$_2$-photodissociating and ionizing radiation from stellar
populations is followed using the \moray\ radiation transport solver
\citep{2011MNRAS.414.3458W}.  The properties of hydrogen and helium
are calculated using a 9-species primordial non-equilibrium chemistry
and cooling network \citep{Abel97}, supplemented by metal-dependent
cooling tables \citep{2009ApJ...691..441S}.  Prescriptions for
Population III and metal-enriched star formation and feedback are
employed, using the same density and metallicity criteria as
\citet{2014MNRAS.442.2560W} but with a Population III characteristic
mass of $40 \Ms$. 

We simulated a region of the universe 28.4 Mpc/h on a side 
using the WMAP7 best-fit cosmology.  Initial conditions were
generated at $z=99$ using MUSIC \citep{2011MNRAS.415.2101H}, and a
low-resolution simulation was run to $z=6$ to find
regions suitable for re-simulation.  The volume was then
smoothed on a physical scale of 5 comoving Mpc, and regions of high
($\langle\delta\rangle \equiv \langle\rho\rangle/(\Omega_M \rho_C) -1
\simeq 0.68$), average ($\langle\delta\rangle \simeq 0.09$), and low
($\langle\delta\rangle \simeq -0.26$) mean density 
were chosen for re-simulation.  These
subvolumes, designated the ``\textit{Rare peak,}''
``\textit{Normal},'' and ``\textit{Void}'' regions, with comoving
volumes of 133.6, 220.5, and 220.5 Mpc$^3$, were resimulated with
an effective
initial resolution of $4096^3$ grid cells and particles in the region
of interest, giving a dark matter mass resolution of $2.9
\times 10^4$~M$_\odot$.  We allowed further
refinement based on
baryon or dark matter overdensity for up to 12 total levels of
refinement (i.e., a maximum comoving resolution of 19 pc).
For more simulation details, see
\citet{2013ApJ...773...83X,2014ApJ...791..110X,2014ApJ...795..144C}.
These simulations were evolved to $z= (15, 12.5, \mathrm{and}~8)$ for
the \textit{Rare peak}, \textit{Normal}, and \textit{Void}
simulations.  In these simulations, the halo mass function is
well-resolved down to $\simeq 2 \times 10^6$~M$_\odot$ ($\simeq
70$~particles/halo), and the simulations contained $(822, 758, 458)$
galaxies having at least 1,000 particles
(M$_{\mathrm{vir}}$~$\simeq 2.9 \times 10^7$~M$_\odot$).  The
simulations were analyzed using the \yt\ analysis tool \citep{2011ApJS..192....9T}.

\section{Results}
\label{sec:results}

\begin{figure}[t] 
   \includegraphics[width=\columnwidth, clip=true]{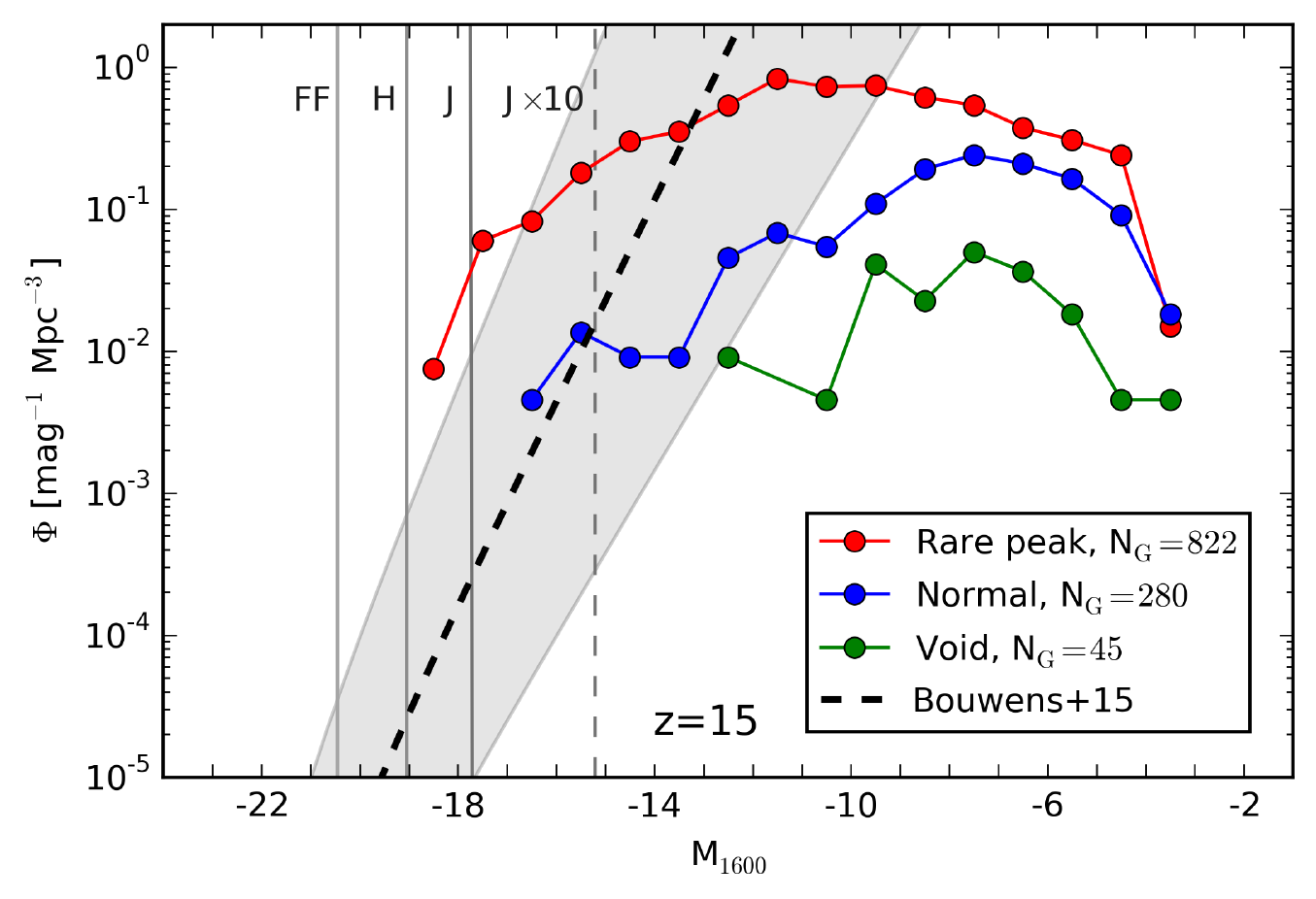} 
   \includegraphics[width=\columnwidth, clip=true]{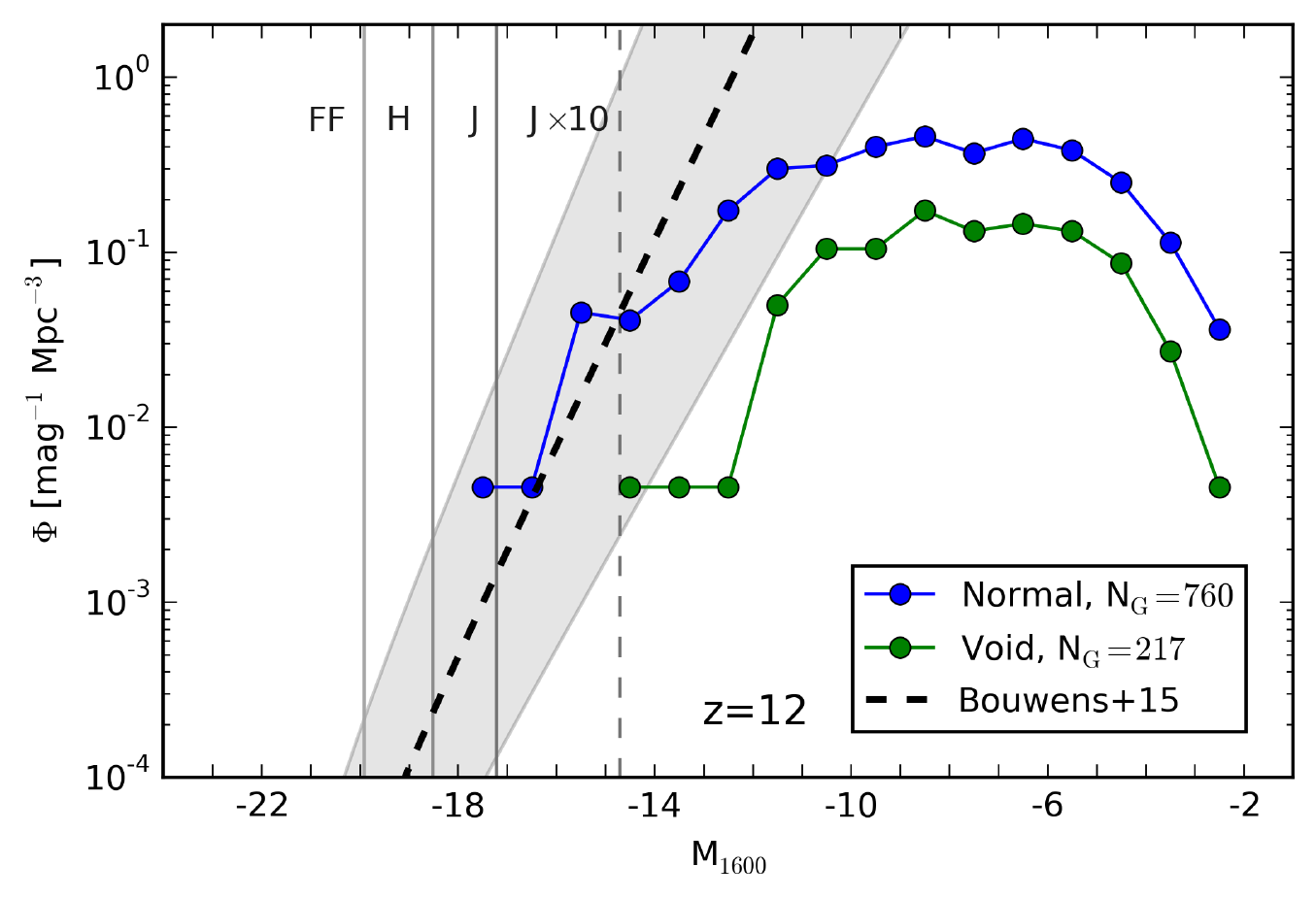} 
   \includegraphics[width=\columnwidth, clip=true]{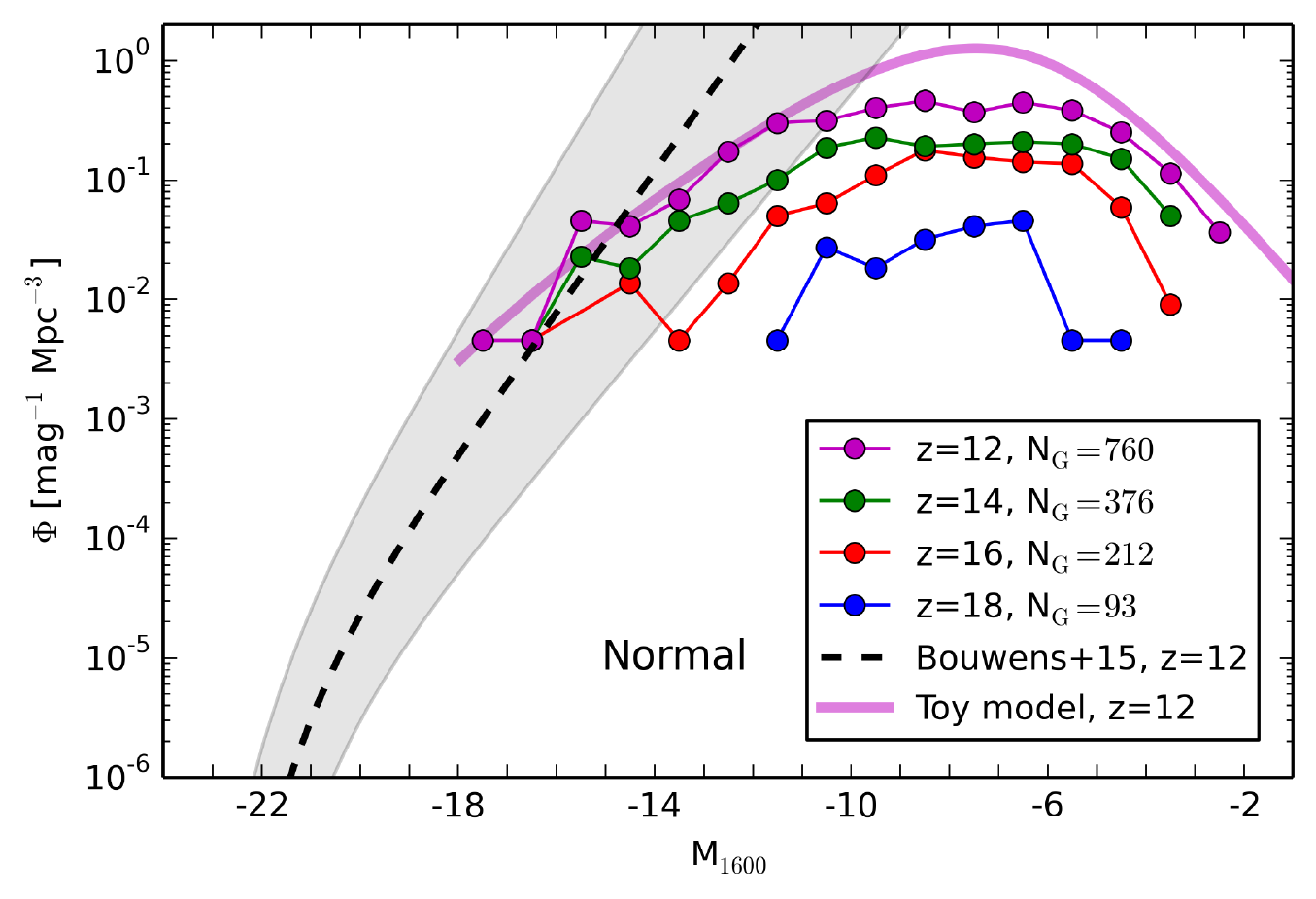} 
   \centering
   \caption{UV luminosity function for galaxies in the Renaissance
Simulations for the \textit{Rare peak}, \textit{Normal}, and
\textit{Void} simulations.  The x-axis shows the absolute UV magnitude
(M$_{1600}$) and the y-axis shows galaxies per magnitude per comoving
Mpc$^3$.  Top row: all simulations at $z=15$.  Middle row:
\textit{Normal} and \textit{Void} simulations at $z=12$.  Bottom row:
\textit{Normal} simulation at $z=18, 16, 14,$ and $12$.  In the top two
panels, the redshift-dependent fit to the Schechter function from
\citet{2015ApJ...803...34B} is shown at the appropriate redshift
($z=12$ for the bottom panel), with the shaded region representing the
$1-\sigma$ uncertainty in the fit parameters.  In each panel, solid vertical grey lines
delineate the detection limits at that redshift for galaxies at the epoch shown
using (from left to right) the limiting magnitudes of the un-lensed
Hubble Frontier Fields (m$_\mathrm{lim} = 28.7$), the HUDF12
(m$_{\mathrm{lim}} = 30.1$), and the JWST ultradeep campaign
(m$_{\mathrm{lim}} = 31.4$).  The vertical dashed-grey line
corresponds to a JWST ultradeep observation with a lensing
magnitification of 10. The magenta line in the bottom panel is a toy
model of the UV luminosity function for this simulation at $z=12$, described in
Section~\ref{sec:conclusions}.}
\vspace{3em}
\label{fig:lumfctn}
\end{figure}

Figure~\ref{fig:lumfctn} shows the ultraviolet (UV) luminosity
function for the galaxies within the refined region of our
simulations.  This luminosity function is calculated by extracting the
mass, formation time, and metallicity of all star particles in each
simulated galaxy at the final timestep of each simulation, and using
the stellar population synthesis models of \citet{BC03} to create a
spectrum for that galaxy (with each star particle representing a star
cluster with a Chabrier mass function).  The UV luminosity in a window
of $\Delta \lambda = 100$ \AA\ centered on $\lambda = 1600$ \AA\ is
extracted from the resulting spectra.  In each panel, the
redshift-dependent fit to the Schechter function from
\citet{2015ApJ...803...34B} is shown at the same redshift as the
simulation data, with the shaded region representing the $1-\sigma$
uncertainty in their fit parameters.  The figure shows the luminosity
function derived for all three simulations at $z=15$, the
\textit{Normal} and \textit{Void} simulations at $z=12$, and the
redshift evolution of the \textit{Normal} simulation from $z=18$ to
$z=12$ (212 to 375 Myr).  In each panel, we have calculated the
magnitude of the faintest galaxy that could in principle be detected
\textit{at that redshift} in an un-lensed Hubble Frontier Field, the
Hubble Ultra-Deep Field, and the planned JWST ultradeep field, as well
as the limiting magnitude for a JWST observation assuming a field with
a magnification factor of $\mu = 10$.  For all observational estimates
we assume a uniform K-correction of $-2$ magnitudes.  In all
simulations the UV luminosity function matches observations reasonably
well at the bright end, but there is a flattening of the luminosity
function in dim galaxies -- galaxies that currently cannot be
observed, but which in the future may be probed by JWST.  We also see
that the UV luminosity function has a termination point, with no
galaxies dimmer than M$_\mathrm{UV} \simeq -2$.  This behavior is
captured reasonably well by a simple toy model, shown in the bottom
panel and described in 
Section~\ref{sec:conclusions}.

\begin{figure*}[]
   \includegraphics[width=.43\textwidth, clip=true]{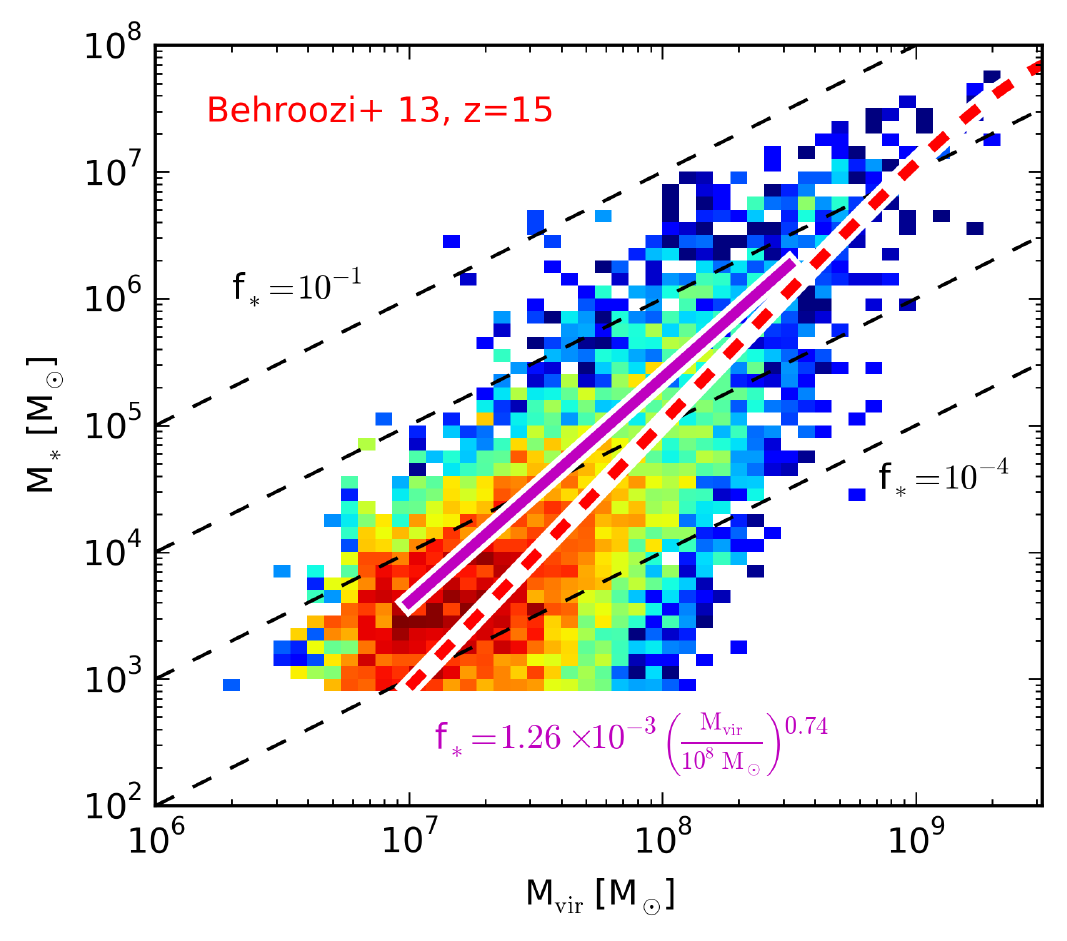} 
   \includegraphics[width=.53\textwidth, clip=true]{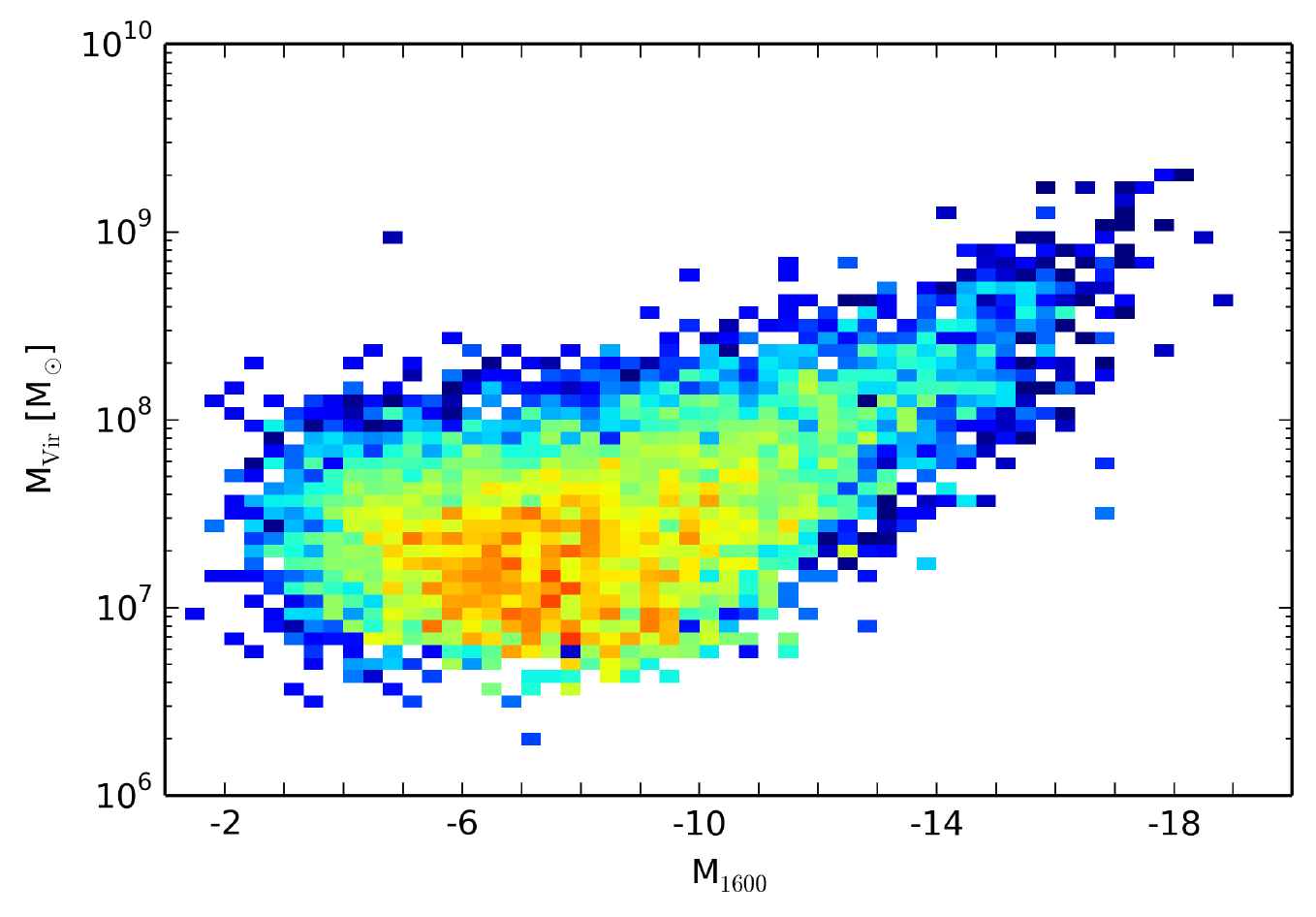} 
   \centering
   \caption{Left panel: 2D histogram of galaxy stellar mass versus
halo virial mass for all simulations combined at their final redshifts.  The histogram is
logarithmic, and the colors indicate the number of galaxies in each
bin (with red/yellow indicating a large number of galaxies and
green/blue indicating few galaxies).  Black-dashed lines indicate
constant stellar fractions (f$_* \equiv$~M$_*$/M$_{\mathrm{vir}}$),
ranging by factors of 10 from f$_* = 10^{-1}$ (top) to $10^{-4}$
(bottom).  The purple solid line and accompanying equation indicates
the fit to the median stellar fraction as a function of virial mass
for halos with $10^7 <$~M$_{\mathrm{vir}}$/M$_\odot < 10^{8.5}$.  The
red-dashed line indicates the stellar fraction as a function of virial
mass predicted by \citet{Behroozi13} but extrapolated to $z=15$.
Right panel: Halo virial mass versus absolute UV magnitude for the
same set of
galaxies.}
   \label{fig:starfraction}
\end{figure*}

Figure~\ref{fig:starfraction} displays the relationship between a
given galaxy's stellar mass and the halo's virial mass or ultraviolet
luminosity, combined at their final redshift.  In the left panel, a 2D histogram of stellar mass versus
halo virial mass, lines of constant stellar mass fraction (defined as
f$_* \equiv$~M$_*$/M$_{\mathrm{vir}}$), ranging from
f$_* = 10^{-1}$ to $10^{-4}$, are plotted.  The purple
solid line and accompanying equation is a fit to the median
stellar fraction as a function of virial mass,

\begin{equation}
f_* = 1.26 \times 10^{-3} \left( \frac{\mathrm{M}_{\mathrm{vir}}}{10^8 \mathrm{M}_\odot}\right)^{0.74}
\end{equation}

\noindent
which is made by fitting a line to the distribution of log stellar
mass and log virial mass for halos with $10^7
<$~M$_{\mathrm{vir}}$/M$_\odot < 10^{8.5}$.  At virial masses larger
than $10^{8.5}$~M$_\odot$, there is some deviation from this
relationship, with the stellar fraction in the most massive halos
  being lower than this power-law fit would predict.  The red-dashed line indicates the stellar fraction as a
function of virial mass predicted by \citet{Behroozi13} but
extrapolated to $z=15$ -- a fit that agrees with the simulated
galaxies to a remarkable degree, including the flattening of the
  stellar fraction at high mass.  The right panel shows a 2D histogram
of halo virial mass versus absolute UV magnitude for the same
galaxies, and indicates a positive (albeit noisy) correlation between halo mass and
star formation rate.
The quantities in both panels are insensitive to redshift during the
epochs considered by this work.

\begin{figure}[] 
   \includegraphics[width=.43\textwidth, clip=true]{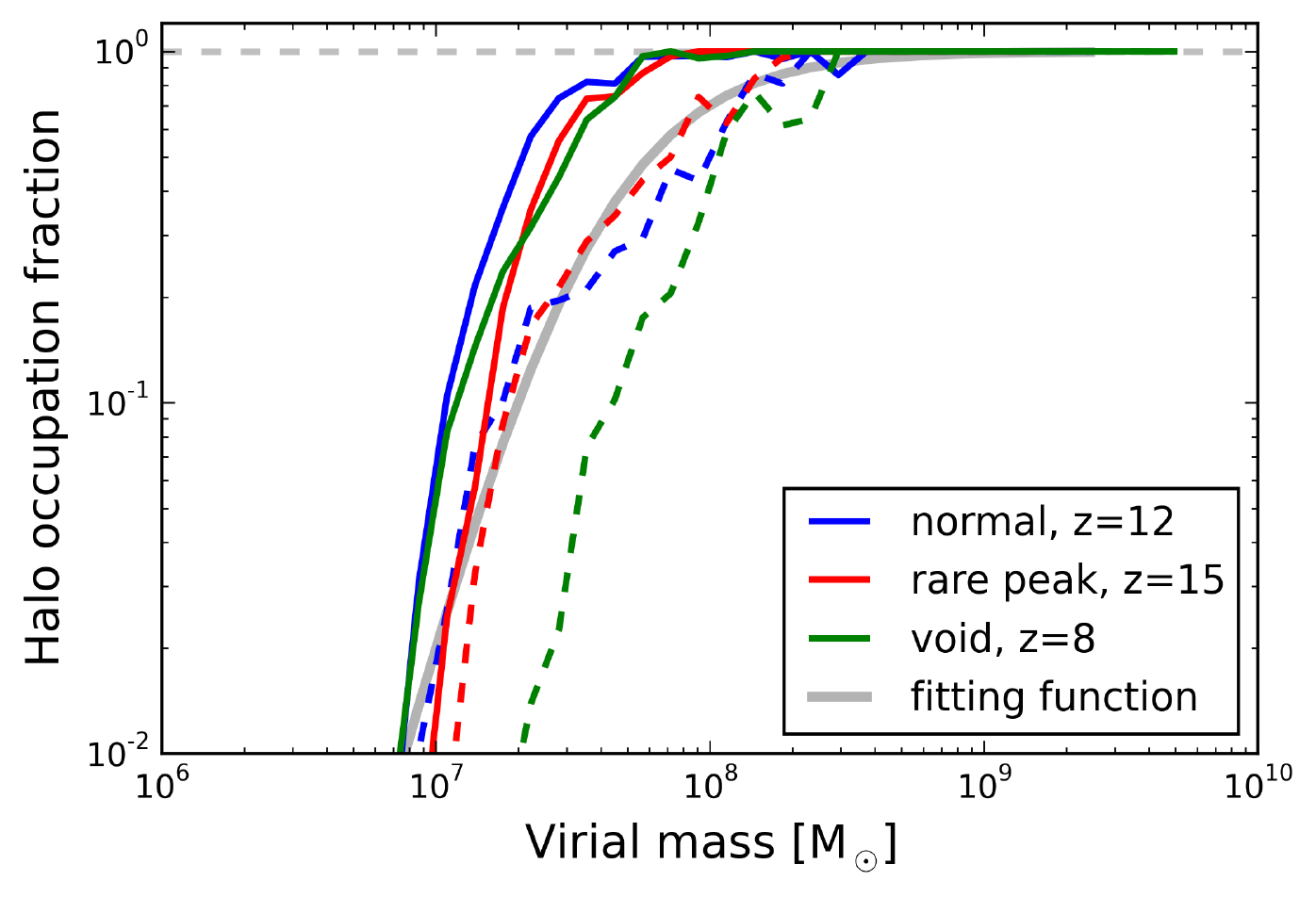} 
   \centering
   \caption{Fraction of halos containing stellar populations as a
function of halo mass for all simulations at their final redshift.
Solid lines: fraction of halos of that virial mass containing metal-enriched stars of any age.  Dashed lines:
fraction of halos of that virial mass that have experienced star formation in the past 20 Myr.  Grey
solid line: fitting function used in the toy model shown in
Figure~\ref{fig:lumfctn}.}
   \label{fig:halofraction}
\end{figure}

Figure~\ref{fig:halofraction} shows the fraction of halos containing
metal-enriched stellar populations as a function of halo virial mass for all three
simulations at their final redshift.  All halos with M$_{\mathrm{vir}}
\ga 5 \times 10^7$~M$_\odot$ contain stars, and all halos with
M$_{\mathrm{vir}} \ga 2 \times 10^8$~M$_\odot$ have formed stars in
the last 20 Myr (and thus are emitting significant ultraviolet light).  No halos
with virial masses below $\simeq 7 \times 10^6$~M$_\odot$ contain 
stars.  This figure also contains a fitting function for the fraction
of halos that have active stars \citep[modeled on the form used by][]{2008MNRAS.390..920O}:

\begin{equation}
f_{\rm occ}(M) = \left[ 1 + \left(2^{\alpha/3} - 1\right) \left(
    \frac{M}{M_c}\right)^{-\alpha} \right]^{-3/\alpha}
\end{equation}

with $M_c = 6.0 \times 10^7$~M$_\odot$ and $\alpha = 1.5$.

\begin{figure*}[] 
   \includegraphics[width=1.0\textwidth, clip=true]{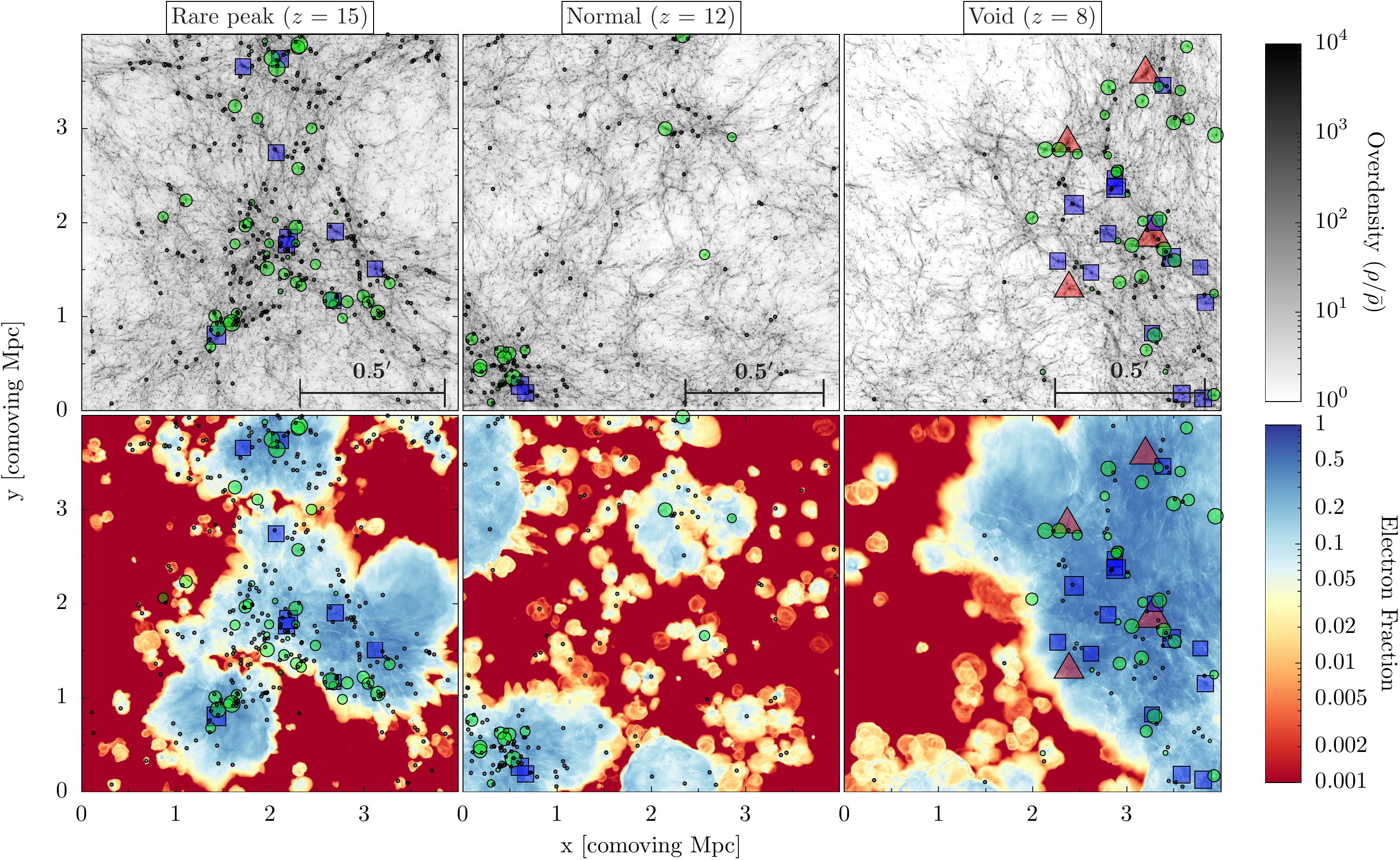} 
   \centering
   \caption{Projections through the refined subvolumes of the 
     \textit{Rare peak}, \textit{Normal}, and \textit{Void}
     simulations at their final redshifts.  Top row: matter
     overdensity.  Bottom row: electron fraction weighted by gas
     overdensity.  In each panel in the top row, the black bar 
     indicates the physical size subtended by a
     field of view $0.5^\prime$ at that redshift, and the
     projection area is slightly smaller than the $2.2^\prime \times
     2.2^\prime$ JWST NIRCam field of view.  In both rows, all
     galaxies with $M_{1600} \le -10$ are
     shown, with those that in principle could be observed by the
     HUDF12 shown as red triangles, by the unlensed JWST ultradeep
     field as blue squares, and considering a magnification factor of
     $\mu = 10$ in the JWST ultradeep field as green circles, with
     glyph size proportional to luminosity.  Note that any galaxy
     observable in the HUDF could also be seen in the JWST ultradeep
     field.  Galaxies with $M_{1600} \le -10$ and unobservable by any
     of these campaigns are shown as black dots.}
   \label{fig:projections}
\vspace{7mm}
\end{figure*}

Figure~\ref{fig:projections} shows projections of matter overdensity
and electron fraction in the refined simulation subvolumes 
at their final redshifts.  In each panel of the top row,
the black bar indicates the physical size
subtended by a field of view $0.5^\prime$ on a side at that redshift.
The projection area is slightly smaller than the $2.2^\prime
\times 2.2^\prime$ JWST NIRCam field of view.  In both rows, all
galaxies with $M_{1600} < -10$ are shown, with glyphs showing galaxies
which could be seen in the 2012 Hubble Ultra Deep
Field (red triangles) \citep{2013ApJS..209....3K}, the un-lensed (blue
squares) and lensed (green circles; $\mu = 10$) JWST ultradeep field
\citep{2006SSRv..123..485G}, and dimmer galaxies (black dots).  The
spatial distribution of galaxies in each field demonstrates cosmic
variance.  The rare peak field shows many clustered galaxies above the
lensed JWST limit, whereas the normal field has very few galaxies
outside of a clustered region.  The void region is bounded on the right by an overdense
sheet containing several galaxies above the JWST unlensed sensitivity
limit.  In the void, there is little collapsed structure that can
sustain efficient star formation, leading
to a lack of galaxies with $M_{1600} < -10$.  The projected electron
fraction in the lower row depicts the \hii~regions with radii up to
$\sim$1 comoving Mpc associated with these galaxies.
Scattered around the volume are relic \hii~regions created by
Population III stars that form in substantially smaller halos ($M_{\rm
  vir} \sim 10^6 \Ms$) than the galaxies shown in the Figure,
marginally adding to the overall ionized fraction.  The overall
ionized morphology is consistent with an inside-out scenario,
previously noted by several groups \citep[e.g.][]{Gnedin00, Trac08,
  Finlator09,2014ApJ...789..149S} during the initial phases of
reionization.  This is most apparent in the void simulation, because the ionization front from the overdense sheet has not yet
propagated into that region and there are no void galaxies to reionize
their surroundings.

\section{Summary and Discussion}
\label{sec:conclusions}

The key results presented in this paper are as follows:

\begin{enumerate}

\item The ultraviolet luminosity function of our simulated galaxies is
consistent with observations of high-z galaxy populations at the
bright end of the luminosity function (M$_{1600} \la -17$), but at
lower luminosities is essentially flat rather than rising steeply as
inferred by a Schechter function fit to observational data, and has a
clearly-defined lower limit in UV luminosity.

\item This behavior of the UV luminosity function is due to two
factors: (i) the strong dependence of the star formation rate on halo
virial mass in our simulated galaxy population, with lower-mass halos
having systematically lower star formation rates and thus lower UV
luminosities; and; and (ii) the fact that halos with virial masses
below $\simeq 2 \times 10^8$~M$_\odot$ do not universally contain
stars, with the fraction of halos containing stars falling with
decreasing virial mass and reaching zero at $\simeq 7 \times
10^6$~M$_\odot$.

\item The brightest of our simulated galaxies may be visible to
current and future ultra-deep space-based surveys at $z \sim 12$ (and
likely at lower redshifts as well),
particularly if lensed regions are chosen for observation.

\end{enumerate}

The primary result of this paper -- the flattening of the mass
function -- is significant for our understanding of the reionization
of the universe.  Observations of high redshift galaxies provide poor
constraints on the low-luminosity end of the galaxy luminosity
function, and thus make it challenging to accurately account for the
full budget of ionizing photons during that epoch.  Our work suggests
that there are far fewer faint galaxies at high redshift than one would infer from
fitting a Schechter function to the high-luminosity end, as has been
done by, e.g., \citet{2015ApJ...803...34B}.  Taken at face
value, a smaller number of galaxies would result in an
\textit{under}-production of photons relative to what is needed to
reionize the universe.  However, the \textit{Planck} 2015 results
\citep{Planck15_Cosmo}, particularly the Thomson scattering optical
depth $\tau = 0.066 \pm 0.012$ corresponding to an instantaneous
reionization redshift $z_r = 8.8_{-1.1}^{+1.2}$, relieve the previous
tension between reionization constraints from polarization
measurements of the CMB and galaxy
observations \citep{Robertson15}.

There are many complicating factors when computing a reionization
history from a galaxy luminosity function -- in particular, the UV
photon escape fraction and its relationship to the halo virial mass
\citep[see, e.g.,][]{2014MNRAS.442.2560W, 2014MNRAS.438.2097F} -- and
we defer a detailed analysis to future work (Xu et al. 2015, in
prep.).  Recent work has shown that low-mass galaxies ($M_{\rm vir}
\la 10^8 \Ms$; $M_\star \la 10^5 \Ms$) prior to reionization have
typical escape fractions in the range 5--50\% with the upper range
occurring when supernova blastwaves and ionization fronts create
low-density channels that allow ionizing radiation to freely escape
into the neutral intergalactic medium \citep{Paardekooper13,
  Paardekooper15, 2014MNRAS.442.2560W}.  In more massive halos, the
simulations of \citet{Kimm14_fesc} showed that the time-averaged UV
escape fraction for a particular galaxy is only $\sim$10\% 
(although \citet{2015arXiv150307880M} suggest that this fraction may
be considerably lower over halos with a mass range of $10^9-10^{11}$~M$_\odot$,
with a time-averaged value of around $\simeq 5\%$).  This
suggests that low-luminosity galaxies may provide a non-negligible
amount of the ionizing photon budget, given their higher escape
fractions and number densities.

The flattening of the high redshift galaxy luminosity function at the
faint end is a robust result\footnote{See Supplementary Online
  Materials.} in our simulations \citep[and has been suggested by
other, albeit less physics-rich, simulations -- see
][]{2013ApJ...766...94J}.  What is less robust, however, is what can
be inferred from the overall normalization of the luminosity function
from our calculations.  Given JWST's small field of view, cosmic
variance will present a significant challenge
\citep[e.g.,][]{2008ApJ...676..767T}.  We have sampled regions of
varying mean density, which gives some sense of cosmic variance.
However, a precise discussion of this subject is difficult because of
the small cosmological volumes sampled, as well as the somewhat
arbitrary nature of our choice of resimulated regions.

We note that the strongly evolving stellar mass fraction as a function
of halo mass has an
intriguing similarity with observations of Local Group dwarf galaxies
\citep[e.g.,][]{2012AJ....144....4M}, where there is a strong positive
correlation between a given galaxy's mass-to-light ratio and its
bolometric luminosity.  The smallest local
dwarf galaxies are strongly dark matter-dominated, having similar
stellar fractions to our lower-mass galaxies, and the trends are similar as
well.  While it is hard to make precise comparisons between Local
Group observations and high-redshift galaxy simulations, the 
similarity in trends suggests that there may be universal, halo
mass-dependent behaviors in galaxy formation.

Figure~\ref{fig:halofraction} indicates something intriguing --
namely, that there is some physical effect that quenches star
formation in low mass halos in the early universe.  The cause is not
reionization -- Figure~\ref{fig:projections} demonstrates that the
universe is patchily reionized at this epoch -- but rather H$_2$-photodissociating
Lyman-Werner radiation \citep[e.g.,][]{2008ApJ...673...14O}.  The
universe is transparent to this radiation, and thus star formation
anywhere in the volume can suppress molecular hydrogen elsewhere, with
the level of suppression growing with the comoving star formation rate
density.  The consequence of this suppression is that low-mass halos of
primordial composition, which cool primarily by H$_2$ line emission,
must grow to be more massive and hotter before the gas in the halo
core can cool efficiently, which suppresses Population III star
formation (and, thus, metal-enriched star formation).  Once halos
reach this threshold mass \citep[which depends on the strengh of the
Lyman-Werner background; see][]{2008ApJ...673...14O}, stars can then
form.  We note that at later times, when the universe is transparent
to hydrogen-ionizing photons, a similar but more pronounced effect
must exist due to this photon population.

Figures~\ref{fig:starfraction} and~\ref{fig:halofraction} strongly
suggest that the flattening of the galaxy luminosity function is due to
a combination of decreased star formation rate at low halo virial masses
and the lack
of stars in halos below $\sim 10^8$~M$_\odot$.  A simple toy
model\footnote{See Supplementary Online Materials for a detailed description.} 
using the analytic halo mass function convolved with a constant
specific star formation rate and the halo stellar
occupation fraction from Figure~\ref{fig:halofraction} is shown as the
magenta line in the bottom panel of Figure~\ref{fig:lumfctn},
and matches the properties of the luminosity function of our simulated
galaxies quite well, accurately capturing the deviation from a
Schechter function, the overall flattening at intermediate luminosity,
and the termination at the low-luminosity end.

The dependence of stellar fraction on halo mass is well-known
observationally \citep[e.g.,][]{Behroozi13}, and is also seen in
simulations of more massive galaxies at lower redshifts
\citep[e.g.,][]{2014MNRAS.445..581H}.  We speculate that this behavior is due to a combination of
cooling and stellar feedback \citep[as has been discussed in a 
different context by][]{,2014arXiv1409.1598V}.  Gas in low-mass halos cools
inefficiently due to the low virial temperature, and stellar
feedack can efficiently remove metal-enriched gas from these low-mass
halos \citep[see, e.g.,][]{2014ApJ...795..144C}, which further
decreases the efficiency with which gas can cool and form stars.  The
suppression of star formation in halos below $\sim 10^8$~M$_\odot$
(and complete absence below $\sim 10^7$~M$_\odot$) is likely due to
inefficient cooling, destruction of H$_2$ via Lyman Werner radiation
from neighboring halos, and the resulting Jeans screening
\citep[e.g.,][]{1998MNRAS.296...44G,2008ApJ...684....1W,2014MNRAS.442.2560W}.  This
suggests that there is an effective lower limit on the masses of
high-redshift galaxies, and thus a definitive lower end to the UV
luminosity function.  We will explore this in future work.

\acknowledgements

This research is part of the Blue Waters project using NSF PRAC
OCI-0832662.  This research was supported by NSF and NASA grants
PHY-0941373, AST-1109243, AST-1211626, AST-1333360, NNX12AC98G,
HST-AR-13261.01-A, HST-AR-13895.001.  The authors thank Devin Silvia
and Eric Bell for useful discussions.

\newpage



\appendix
\section{Supplementary Online Material}
\label{app:som}

\subsection{A Toy UV Luminosity Function Model}

The simple luminosity function model shown by the magenta line in the
bottom panel of Figure~\ref{fig:lumfctn} was created using the
following procedure:

\begin{enumerate}

\item Calculate the comoving number density of cosmological halos as a
function of halo virial mass m, $\frac{dN(m)}{dV}$, at $z=12$ for the WMAP7 best-fit
cosmology, using the fitting function of \citet{2006ApJ...646..881W}.

\item For each halo mass bin, calculate the recent star formation
history of the halo assuming a constant specific star formation rate
of $10^{-8}$~yr$^{-1}$ over the last 100 Myr.  This results in a
constant star formation rate, under the assumption of
constant halo mass.

\item Given this star formation history and the assumption of a simple
Salpeter IMF with parameters identical to those in the $z=0$ Milky Way
(consistent with our simulation treatment of metal-enriched stars; see
Section~\ref{sec:methods}), calculate the IMF-integrated bolometric
luminosity of the stars that have not yet undergone supernovae at the
epoch under consideration.  Given that this luminosity is dominated by
the most massive and hottest stars (since luminosity scales
approximately as stellar mass to the fourth power), we make the
assumption that the UV luminosity of the stellar population in each
halo is roughly its bolometric luminosity.  This results in a comoving
number density of halos as a function of UV magnitude,
$\frac{dN}{dVdM}$, where M represents absolute UV magnitude.

\item Finally, we multiply the resulting quantity by the fraction of
halos that have experienced recent star formation as a function of
halo mass that is shown in Figure~\ref{fig:halofraction}, using the
crude approximation shown by the grey line.  This fit captures the
essence of the figure -- namely, universal recent star formation above
M$_{\rm vir} \geq 2 \times 10^8$~M$_\odot$, no recent star formation below
$\sim 10^7$~M$_\odot$, and a steady decrease in star formation with
decreasing halo mass between those endpoints.

\end{enumerate}

This toy model captures several of the essential features of the
luminosity function shown by our simulations: it has a significant
slope at the high-luminosity end, flattens out at $M_{1600} > -12$,
and goes entirely to zero at $M_{1600} \simeq -2$.  Furthermore, the
properties of this toy model are robust to several of the simplifying
assumptions made here.  The most critical of these are as follows:

\noindent
\textbf{Specific star formation rate:} The specific star formation
rate (SSFR) evolves as both a function of redshift and halo mass.
This evolution is most pronounced at low redshift ($z < 2$)
-- at higher redshifts, such as those considered in this work, the
SSFR evolves very slowly with redshift and is relatively
insensitive to halo mass \citep[see, e.g., Appendix F
of][]{Behroozi13}, making the approximation of a constant SSFR
sensible.  We note that at the redshifts and halo masses of interest
to the work presented here this quantity is extremely hard to
constrain observationally; however, extrapolating the
observed trends to $z > 8$ suggests little evolution.  Deviation from
a constant SSFR will modify the predicted UV luminosity function in
proportion to the deviation.

\noindent
\textbf{Halo formation history:} Cosmological halos grow over time,
and do so particularly quickly at the redshifts considered in this
work.  Assuming that the specific star formation rate is directly
proportional to halo mass, this implies that the total (as opposed to
specific) star formation rate in a given halo is increasing rapidly as
well.  The period of 100 Myr over which the star formation rate is
assumed to be constant is a significant fraction of the Hubble time at
high redshift ($\simeq 250$~Myr at $z=16$, $\simeq 380$~Myr at
$z=12$), suggesting that this assumption may be suspect.  However, the
UV luminosity is dominated by massive stars having lifetimes of
millions of years, and thus the production of UV photons is dominated
almost entirely by the very recent star formation history of the halo.
As a result, the UV luminosity is quite robust to the exact details of
the halo's formation history, with deviations from constant star
formation rate contributing mildly to variation in magnitude.

\subsection{Comparison with Higher Resolution Simulations}

The simulated behavior of the UV luminosity function at the faint-end
can be suspectible to limited numerical resolution.  To check for any
spurious effects, we compare our results to our previous work
\citep{2012ApJ...745...50W, 2012MNRAS.427..311W, 2014MNRAS.442.2560W}
that focused on 32 first-generation galaxies in a small comoving
volume of (1 Mpc)$^3$ with a DM particle mass resolution of 1840~\Ms,
15 times smaller than the dark matter particles in the simulations presented in this paper.  The major
shortcoming of our previous work was the limited galaxy sample size,
which was rectified in the Renaissance Simulations.  These two
simulation suites share the same subgrid physics models --- chemical
reaction networks, metal-free and metal-enriched star formation, and
radiative and supernova feedback.  Although the Renaissance
Simulations cannot capture Population III star formation in the smallest
minihalos ($\sim 2 \times 10^5 \Ms$), properties of metal-enriched
star formation, in particular the stellar mass -- halo mass relation,
agree very well between the two simulation suites \citep[see Fig. 4
in][]{2014ApJ...795..144C}.  There we showed that this relationship in
both simulations were within 1-$\sigma$ of each other in halos with
$M_{\rm vir} \ge 10^7 \Ms$, which are the smallest halos considered in
our analysis.  Furthermore, \citet{2014MNRAS.442.2560W} found that the
UV luminosity function was flat at magnitudes $M_{1600} \ga -12$,
consistent with our findings in the Renaissance Simulations.  Thus, we
are confident that the flattening of the luminosity function is not a
result of limited numerical resolution but is caused by the
suppression of star formation from radiative and supernova feedback
effects in halos with masses $M_{\rm vir} \la 2 \times 10^8 \Ms$.

\end{document}